\documentclass{JHEP3}
\usepackage{amssymb,amsmath}
\usepackage{epsfig}
\usepackage{graphicx}
\usepackage{citesort}

\title{On the role of shear in cosmological averaging II: large voids, non-empty voids and a network of different voids}

\author{Maria Mattsson$^{1,2,}$\footnote{E-mail: maria.ronkainen@helsinki.fi}~, Teppo Mattsson$^{2,}$\footnote{E-mail: teppo.mattsson@canterbury.ac.nz}\\
${}^1$ Physics Department and Helsinki Institute of Physics, P.O.Box 64, FIN-00014 University of Helsinki, Finland\\
${}^2$ Department of Physics and Astronomy, University of
Canterbury, Private Bag 4800, Christchurch 8140, New Zealand\\}

\abstract{We study the effect of shear on the cosmological
backreaction in the context of matching voids and walls together
using the exact inhomogeneous Lema\^itre-Tolman-Bondi solution.
Generalizing JCAP {\bf 1010} (2010) 021, we allow the size of the
voids to be arbitrary and the densities of the voids and walls to
vary in the range $0 \leq \Omega_{\rm v} \leq \Omega_{\rm w} \leq
1$. We derive the exact analytic result for the backreaction and
consider its series expansion in powers of the ratio of the void
size to the horizon size, $r_0/t_0$. In addition, we deduce a very
simple fitting formula for the backreaction with error less than
$1\%$ for voids up to sizes $r_0 \gtrsim t_0$. We also construct
an exact solution for a network of voids with different sizes and
densities, leading to a non-zero relative variance of the
expansion rate between the voids. While the leading order term of
the backreaction for a single void-wall pair is of order
$(r_0/t_0)^2$, the relative variance between the different voids
in the network is found to be of order $(r_0/t_0)^4$ and thus very
small for voids of the observed size. Furthermore, we show that
even for very large voids, the backreaction is suppressed by an
order of magnitude relative to the estimate obtained by treating
the walls and voids as disjoint Friedmann solutions. Whether the
suppression of the backreaction due to the shear is just a
consequence of the restrictions of the used exact models, or a
generic feature, has to be addressed with more sophisticated
solutions.}

\preprint{HIP-2010-38/TH}

\keywords{Inhomogeneous Cosmological Models, Averaging in General
Relativity, Cosmology, Gravitation}

\begin{document}

\section{Introduction}\label{intro}

The non-commutativity of time evolution and spatial averaging in
general relativity implies that the average expansion of a
universe with structures does not evolve in time like the uniform
Hubble expansion in a homogeneous Friedmann model
\cite{Ell84,EllisStoeger,Ellis:2005uz}. This effect can be
quantified by a backreaction term in generalized Friedmann
equations and understood as a consequence of the nonlinearity of
gravity \cite{Buchert:1999er}. Although the cosmological
backreaction is conceptually well-understood, the complexity of
the structure formation at the nonlinear level means its magnitude
in the real universe is difficult to evaluate and is hence widely
debated \cite{Ellis:2008zza}: some studies have found that the
backreaction is small
\cite{Ishibashi:2005sj,Paranjape:2008jc,Clifton:2009jw,Clifton:2010fr,Alonso:2010zv,Green:2010qy},
while other studies suggest it can have a significant effect on
the cosmological dynamics and observations
\cite{Rasanen:2006kp,Wiltshire:2007jk,Mattsson:2007tj,Rasanen:2008it,Wiegand:2010uh},
even to the extent of accounting for the observed cosmic
acceleration entirely without additional effects
\cite{Rasanen:2006kp,Rasanen:2008it,Wiegand:2010uh}. As many of
the current standard values for the cosmological parameters -- not
just the cosmological constant -- rely on the hypothesis of
negligible backreaction, its evaluation has become a central issue
in cosmology today \cite{Sarkar:2007cx,Blanchard:2010gv}.

A widely used scheme to estimate the backreaction is to average
the scalar parts of the Einstein equation on spacelike
hypersurfaces, defined by constant proper time of the
freely-falling dust particles. In this so-called Buchert approach,
the backreaction term is given by the (positive) variance of the
expansion rate minus the (positive) average shear
\cite{Buchert:1999er}. In studies that have estimated the
backreaction to be significant
\cite{Rasanen:2006kp,Wiltshire:2007jk,Rasanen:2008it,Wiegand:2010uh},
the shear on the boundaries between regions characterized by
different expansion rates has been neglected or, equivalently, the
boundary regions or matching conditions have been ignored. On the
other hand, in perturbative studies that do not neglect the shear
the backreaction has been found to be small
\cite{Brown:2008ra,Clarkson:2009hr,Umeh:2010pr}; however, see
\cite{Rasanen:2010wz} for a discussion on the possible
shortcomings of the perturbation theory in modelling the structure
formation. Considering that on physical grounds, shear is expected
to occur on the boundaries between regions of different expansion
rates, evaluating the effect of the shear on the backreaction is
evidently one of the key issues in the problem.

This work is a continuation to our previous work
\cite{Mattsson:2010vq}, hereafter paper I, in which the effect of
shear on the cosmological backreaction was studied in the context
of matching voids (with $\Omega_{{\rm v}} = 0$ and $H t = 1$) and
walls (with $\Omega_{{\rm w}} = 1$ and $H t = 2/3$) together using
the exact inhomogeneous Lema\^itre-Tolman-Bondi or LTB solution.
Whereas neglecting the exact matching can lead to significant
backreaction, the main conclusion of paper I was that the shear
arising from the exact matching suppresses the backreaction by the
squared ratio of the void size to the horizon size, $(r_0/t_0)^2$,
thus making it small for voids of the observed size $r_0/t_0
\lesssim 10^{-2} $ \cite{Hoyle:2003hc}. Here we generalize the
study of paper I by considering:
\begin{enumerate}
\item The backreaction as a function of the void and wall density
parameters in the range $0 \leq \Omega_{{\rm v}} \leq \Omega_{{\rm
w}} \leq 1$, thus relaxing the priors $\Omega_{{\rm v}}=0$ and
$\Omega_{{\rm w}}=1$ of paper I. \item The backreaction for voids
of arbitrary size $r_0$, thus relaxing the condition $r_0 \ll t_0$
assumed in paper I. \item A network of voids with different
densities $\Omega_{{\rm v}}$ and radii $r_0$, thus giving rise to
relative variance of the expansion rate between the different
void-wall pairs.
\end{enumerate}
Note that since the backreaction of the quasi-spherical Szekeres
model reduces to the LTB model, the results of this work apply to
the quasi-spherical Szekeres solution as well
\cite{Bolejko:2008zv,Bolejko:2010wc}.

Although we focus here on the effect of the structure formation on
the average dynamics of the universe, the ultimate goal of the
research is to evaluate the total effect of the structure
formation on the cosmological observations. In addition to the
dynamical backreaction considered in this work, known effects of
the structure formation include modifications on the propagation
of light not directly determined by the volume-averaged expansion
\cite{Mattsson:2007tj,GarciaBellido:2008nz,GarciaBellido:2008gd,GarciaBellido:2008yq,Rasanen:2008be,Enqvist:2009hn,Clifton:2009jw,Kainulainen:2009dw,Blomqvist:2009ps,Rasanen:2009uw,Krasinski:2010rc,Biswas:2010xm,Clarkson:2010ej,Yoo:2010hi,Bolejko:2010nh,Davis:2010jq,Nadathur:2010zm}
and effects due to our non-average location in the universe
\cite{Wiltshire:2007jk,Wiltshire:2008sg,Wiltshire:2009db,Smale:2010vr}.
These effects may be important but we do not try to evaluate them
in this work.

The paper is organized as follows. The necessary background of the
Buchert averaging method and the LTB solution are introduced in
Sects.\ \ref{averaging} and \ref{ELTEEBEE}, respectively. Sect.\
\ref{model} provides the definition of the void-wall LTB model,
which we apply to study the effect of shear on the backreaction in
Sect.\ \ref{backreactioninLTB}: The general analytic expression of
the backreaction for the void-wall LTB model is derived in Sect.\
\ref{generalQ} and its expansion as a power series and a simple
but accurate fitting formula are considered in Sect.\
\ref{powerexpansionofQ}. The result for the backreaction is then
applied to a network of different voids in Sect.\
\ref{voidnetwork} and to large voids in Sect.\ \ref{largevoids}.
In Sect.\ \ref{kompensoimatonna}, we discuss uncompensated voids.
Finally, the results are summarized in Sect.\ \ref{konkluusiot}
and the conclusions are given in Sect.\ \ref{konkluusiot2}.

\section{Scalar averaging}\label{averaging}

The spatial volume-average of a scalar $S(x,t)$ is defined as
\begin{equation}\label{defAverage}
\langle S \rangle_{\mathcal{D}} (t) \equiv \frac{
\int_{\mathcal{D}}S (x,t) \epsilon(x,t)}{\int_{\mathcal{D}}
\epsilon(x,t)}~,
\end{equation}
where the volume element is determined by the determinant of the
spatial part of the metric as $\epsilon = \sqrt{\det g_{ij}} {\rm
d}^3 x$ and $\mathcal{D}$ is the averaging domain or a region of
the spatial sections. Unless otherwise noted, we take the
averaging domain to be an origin-centered ball of coordinate
radius $R$, i.e.\ $\mathcal{D}=\mathbb{B}(R)$, and simply write $
\langle S \rangle \equiv \langle S \rangle_{\mathbb{B}(R)}$.

The volume expansion scalar $\theta$ is defined in terms of the
temporal change of the volume,
\begin{equation}\label{thetaDef}
\theta (x,t) \equiv \partial_t \ln (\epsilon(x,t)) ~,
\end{equation}
and is related to the generalized scale factor $a(t)$ as
\begin{equation}\label{thetaintermsofscalefactor}
 \langle \theta \rangle (t) = 3 \frac{\dot{a}(t)}{a(t)} ~,
\end{equation}
where
\begin{equation}\label{ExtendedScaleFactor}
a(t)\equiv\left( \frac{\int \epsilon(x,t)}{\int
\epsilon(x,t_0)}\right)^{1/3}~.
\end{equation}
By taking the time derivative of the average (\ref{defAverage})
and writing the expression in terms of the expansion scalar
(\ref{thetaDef}), we obtain the commutator of time evolution and
averaging
\begin{equation}\label{commutator}
\partial_t \langle S \rangle = \langle
\partial_t S \rangle + \langle S \theta \rangle - \langle S
\rangle \langle \theta \rangle ~,
\end{equation}
which, when applied to the expansion scalar itself, yields
\begin{equation}\label{commutatortheta}
\partial_t \langle \theta \rangle = \langle \partial_t
\theta \rangle + \underbrace{\langle (\theta - \langle \theta
\rangle )^2 \rangle}_{\geq 0} ~.
\end{equation}
The result (\ref{commutatortheta}) shows that the average
expansion always decreases less (or increases more) in time than
the average of the time derivative of the local expansion.

By applying the averaging (\ref{defAverage}) to the scalar parts
of the Einstein equation in an irrotational dust universe, and
using the results (\ref{thetaintermsofscalefactor}) and
(\ref{commutator}), we obtain the generalized Friedmann equations
\cite{Buchert:1999er}:
\begin{eqnarray}
3\frac{\ddot{a}(t)}{a(t)} &=& -4 \pi G \langle\rho\rangle(t)+\mathcal{Q}(t)~ \label{BuchertFirst2} \\
3\left(\frac{\dot{a}(t)}{a(t)}\right)^2 &=& 8 \pi
G\langle\rho\rangle(t)-\frac{1}{2}\langle {\mathcal{R}}
\rangle(t)-\frac{1}{2}\mathcal{Q}(t)~
\label{BuchertSecond2} \\
\partial_t\langle\rho\rangle(t)&=&-3\frac{\dot{a}(t)}{a(t)}\langle\rho\rangle(t)~,
\label{BuchertThird2}
\end{eqnarray}
where ${\mathcal{R}}$ is the Ricci curvature scalar of the spatial
sections and the backreaction $\mathcal{Q}$ is given by the
variance of the expansion rate $\theta$ minus the average of the
shear scalar $\sigma^2$,
\begin{equation}\label{defBackreaction}
\mathcal{Q}(t) \equiv \frac{2}{3}\left( \langle \theta^2 \rangle -
\langle \theta \rangle^2 \right) - \langle \sigma^2 \rangle ~.
\end{equation}
The backreaction (\ref{defBackreaction}) quantifies the difference
of the time evolution of the averages relative to the uniform
quantities in a homogeneous Friedmann dust universe.

By partitioning the averaging domain $\mathcal{D}$ into a set of
$N$ mutually disjoint subregions $\mathcal{D}_i$ that satisfy $
\bigcup_{i=1}^{N} \mathcal{D}_i = \mathcal{D} $, the backreaction
(\ref{defBackreaction}) can be written as
\begin{equation}\label{TotalBackreaction}
\mathcal{Q}_{\mathcal{D}}= \sum_{i=1}^{N} f_{i}
\mathcal{Q}_{\mathcal{D}_i} +
\frac{1}{3}\sum_{i=1}^{N}\sum_{j=1}^{N}
f_if_j\left(\langle\theta\rangle_{\mathcal{D}_i}-\langle\theta\rangle_{\mathcal{D}_j}\right)^2~,
\end{equation}
where $f_i$ is the volume fraction of the subregion
$\mathcal{D}_i$, $f_i \equiv {\rm Vol}(\mathcal{D}_i)/{\rm
Vol}(\mathcal{D})$. The expression (\ref{TotalBackreaction}) makes
it explicit that the backreaction is an intensive, rather than
extensive, quantity and becomes useful in Sect.\
\ref{voidnetwork}, where we consider a network of voids
constructed by matching together different LTB solutions.

\section{LTB solution}\label{ELTEEBEE}

The exact spherically symmetric dust solution of general
relativity was discovered by Lema\^itre in 1933
\cite{Lemaitre:1933qe} and is now commonly referred to as the LTB
metric:
\begin{equation}\label{LTBmetric}
{\rm d}s^2 = - {\rm d}t^2 + \frac{[A'(r,t)]^2}{1-k(r)}{\rm d}r^2 +
A^2(r,t) ({\rm d} \theta^2 + \sin^2 \theta {\hspace{1pt}} {\rm d}
\varphi^2)~,
\end{equation}
where the prime stands for the radial derivative $A'(r,t) \equiv
\partial_r A(r,t)$, $k(r)$ is related to the spatial Ricci
curvature scalar as
\begin{equation}\label{3RicciScalar}
{\mathcal{R}} = 2 \frac{\partial_r ( A(r,t) k(r) )}{A^2(r,t)
A'(r,t)} ~,
\end{equation}
and $A(r,t)$ is determined by the Friedmann-like evolution
equation, which, following the notation and parametrization
introduced in Ref.\ \cite{Enqvist:2006cg}, reads as
\begin{equation}\label{LTBfriedman}
H(r,t) = H_0(r) \left[ \Omega_0(r) \left(\frac{A_0(r)}{A(r,t)}
\right)^3 + (1- \Omega_0(r)) \left(\frac{A_0(r)}{A(r,t)} \right)^2
\right]^{1/2}~,
\end{equation}
where $H(r,t) \equiv \partial_t A(r,t)/A(r,t) \equiv
\dot{A}(r,t)/A(r,t)$, $H_0(r) \equiv H(r,t_0)$ and $\Omega_0(r)$
are boundary condition functions specified on a spatial
hypersurface $t=t_0$ that determine the radial inhomogeneity
profile, while the freedom to choose the function $A_0(r) \equiv
A(r,t_0)$ corresponds to the scaling of the $r$-coordinate. The
coordinate freedom is used here to set
\begin{equation}\label{Coordinates}
A_0(r) = r~.
\end{equation}
The curvature function $k(r)$ in the metric (\ref{LTBmetric}) is
related to these by
\begin{equation}\label{kcurvature}
k(r) \equiv H_0^2(r) A_0^2(r) (\Omega_0(r) - 1)
\end{equation}
and the boundary condition function $\Omega_0(r)$ is related to
the physical matter density $\rho(r,t)$ on the $t=t_0$
hypersurface as
\begin{equation}\label{LTBomega}
\Omega_0(r) \equiv \frac{8 \pi G}{3 H_0^2 (r)}
\frac{\int_{\mathbb{B}(r)} \rho_0(r) {\rm d}^3
x}{\int_{\mathbb{B}(r)} {\rm d}^3 x}~,
\end{equation}
where $\rho_0(r) \equiv \rho(r,t_0)$ and ${\rm d}^3 x \equiv r^2
\sin \theta \hspace{1pt} {\rm d} r \hspace{1pt} {\rm d} \theta
\hspace{1pt} {\rm d} \varphi$.

For general functions $\Omega_0(r)$ and $H_0(r)$ that are
independent of each other, the LTB solution contains both decaying
and growing inhomogeneities \cite{Silk}. However, given the
observed near isotropy of the CMB, the models with close to
homogeneous early universe form perhaps the most relevant subcase
of the LTB solutions. This corresponds to solutions where growing
modes dominate so we only consider inhomogeneity profiles that
obey the constraint \cite{Mattsson:2010vq}:
\begin{equation}\label{H0-OmegaRelationAppr}
H_0(r)=\frac{1}{t_0}\left(1-\frac{\sqrt{\Omega_0(r)}}{3}\right)~.
\end{equation}

To calculate the backreaction (\ref{defBackreaction}) for the LTB
solution in Sect.\ \ref{backreactioninLTB}, we need the following
quantities: the shear scalar
\begin{equation}\label{LTBshear1}
\sigma^2(r,t) \equiv\sigma^{\mu\nu}\sigma_{\mu\nu} = \frac{2}{3}
\left( \frac{\dot{A}(r,t)}{A(r,t)} - \frac{\dot{A}'(r,t)}{A'(r,t)}
\right)^2~,
\end{equation}
the volume expansion scalar (\ref{thetaDef})
\begin{equation}\label{LTBexpansion}
\theta(r,t) = 2 \frac{\dot{A}(r,t)}{A(r,t)} +
\frac{\dot{A}'(r,t)}{A'(r,t)} = \frac{\partial_r ( A^2(r,t)
\dot{A}(r,t) )}{A^2(r,t) A'(r,t) } ~,
\end{equation}
and their expressions on the $t=t_0$ hypersurface:
\begin{equation}\label{LTBshear2}
\sigma^2 (r,t_0) = \frac{2}{3}\left(rH'_0(r)\right)^2~,
\end{equation}
\begin{equation}\label{LTBtheta}
\theta(r,t_0) = 3H_0(r)+rH'_0(r)=\frac{1}{r^2}
\partial_r \left(r^3H_0(r)\right)~.
\end{equation}
For volume averaging, we also need the LTB volume element
\begin{equation}\label{LTBvolume}
\epsilon(x,t) = \sqrt{\det g_{ij}} {\hspace{1pt}} {\rm d}r
{\hspace{1pt}} {\rm d}\theta {\hspace{1pt}} {\rm d}\varphi =
\frac{A'(r,t)A^2(r,t) \sin \theta}{\sqrt{1-k(r)}} {\hspace{1pt}}
{\rm d}r {\hspace{1pt}} {\rm d}\theta {\hspace{1pt}} {\rm
d}\varphi~.
\end{equation}

\section{The void-wall LTB model}\label{model}

We consider an LTB solution consisting of two different regions: a
void with density parameter $\Omega_{\rm v}$ and a wall with
density parameter $\Omega_{\rm w}$, such that the boundary
condition function (\ref{LTBomega}) has the form
\begin{equation}\label{LTBomegaspecified}
\Omega_0(r)=\left( \sqrt{\Omega_{\rm v}} + (\sqrt{\Omega_{\rm w}}
- \sqrt{\Omega_{\rm v}}) \Theta (r-r_0) \right)^2~,
\end{equation}
where $\Theta$ stands for the Heaviside step function and $r_0$
determines the size of the void. Because of the constraint
(\ref{H0-OmegaRelationAppr}), the density profile
(\ref{LTBomegaspecified}) implies the expansion profile
\begin{equation}\label{HoLTB}
H_0(r) = \frac{t_0^{-1}}{3} \left( 3 - \sqrt{\Omega_{\rm v}} +
(\sqrt{\Omega_{\rm v}}-\sqrt{\Omega_{\rm w}}) \Theta(r-r_0)
\right)~.
\end{equation}
To calculate the backreaction (\ref{defBackreaction}) for the
void-wall model defined by Eqs.\ (\ref{LTBomegaspecified}) and
(\ref{HoLTB}) in Sect.\ \ref{backreactioninLTB}, we also need the
first derivative of the expansion profile:
\begin{equation}\label{H0prime}
H_0'(r) = -\frac{1}{3}t_0^{-1} (\sqrt{\Omega_{\rm w}} -
\sqrt{\Omega_{\rm v}}) \delta(r-r_0)~,
\end{equation}
where $\delta$ stands for the Dirac delta function.

Note that in the case $\Omega_{{\rm v}}=0$ and $\Omega_{{\rm
w}}=1$ the void-wall profile reduces to the step-function limit $n
\rightarrow \infty$ of the LTB model considered in paper I. The
reason to consider only the step-function profile is that, as
already verified in paper I, the dependence of the backreaction on
the sharpness of the transition between the void and the wall is
weak. The step-function transition has also the advantage of
allowing us to analytically perform calculations that would
otherwise call for approximations or numerical methods.

\section{Backreaction in the void-wall LTB model}\label{backreactioninLTB}

In this section, we calculate the backreaction
(\ref{defBackreaction}) for the void-wall LTB model defined by the
density profile (\ref{LTBomegaspecified}) and the expansion
profile (\ref{HoLTB}). Apart from the profile, the systematic
study of the role of shear makes our approach different from the
previous studies on the backreaction in the LTB model which have
focused on finding profiles that exhibit acceleration of the
average expansion
\cite{Nambu:2005zn,Chuang:2005yi,Paranjape:2006cd,Kai:2006ws,Bolejko:2008yj},
on scale-dependence of the averages \cite{Mattsson:2007qp} or on
general properties of the backreaction
\cite{Sussman:2008xp,Sussman:2008vs}.

When considering numerical values for the backreaction, we give
the results in units of the backreaction obtained by taking the
variance of two disjoint Friedmann models: the empty Milne
solution for the void and the spatially flat Einstein-de-Sitter
solution for the wall, yielding the result found in paper I:
\begin{equation}\label{QFRW}
\mathcal{Q}_{{\rm FRW}} = \frac{1}{6} t_0^{-2} ~,
\end{equation}
where the regions are chosen to have equal volumes. These units
are convenient as they directly address the role of shear by
telling us how much the backreaction is suppressed relative to the
Friedmann estimate which neglects the shear altogether.

\subsection{General expression}\label{generalQ}

With the help of Eqs.\ (\ref{LTBshear1}) and (\ref{LTBexpansion}),
the backreaction (\ref{defBackreaction}) of the LTB solution
simplifies to
\begin{equation}\label{backreaction}
\mathcal{Q}(t) = 2 \left\langle \frac{\dot{A}^2(r,t)}{A^2(r,t)}
\right\rangle + 4 \left\langle
\frac{\dot{A}(r,t)\dot{A}'(r,t)}{A(r,t)A'(r,t)} \right\rangle -
\frac{2}{3} \left\langle 2 \frac{\dot{A}(r,t)}{A(r,t)} +
\frac{\dot{A}'(r,t)}{A'(r,t)} \right\rangle^2 ~,
\end{equation}
which, when evaluated on the $t=t_0$ hypersurface using Eqs.\
(\ref{LTBexpansion}) and (\ref{LTBtheta}), reduces to
\begin{equation}\label{backreaction t0}
\mathcal{Q}(t_0) = 6\left(\left\langle H_0^2 \right\rangle -
\left\langle H_0 \right\rangle^2\right) + 4 \left( \left\langle
rH_0'H_0 \right\rangle- \left\langle rH_0' \right\rangle
\left\langle H_0 \right\rangle \right) - \frac{2}{3}\left\langle
rH_0' \right\rangle^2~,
\end{equation}
where $H_0 \equiv H_0(r)$.

For the LTB solution, we write the volume average
(\ref{defAverage}) as:
\begin{equation}\label{average}
\langle S\rangle = \frac{\int_{\mathbb{B}(R)}
S(r,t)\epsilon(r,t)}{
\int_{\mathbb{B}(R)}\epsilon(r,t)}=\frac{\displaystyle
\int_0^RS\frac{r^2{\rm d}r}{ \sqrt{1-k(r)}}}{\displaystyle
\int_0^R\frac{r^2{\rm d}r}{\sqrt{1-k(r)}}} \equiv \frac{[S]}{v
t_0^3} ~,
\end{equation}
where we have defined the reduced dimensionless volume $v$, which,
after the change of variables $r \equiv t_0 y$, reads as:
\begin{equation}\label{Volume2}
v = \int_0^{\varepsilon x}\frac{y^2{\rm
d}y}{\sqrt{1+y^2\alpha(\Theta)}}~,
\end{equation}
where $\varepsilon \equiv r_0/t_0$, $x \equiv R/r_0>1$, $\Theta
\equiv \Theta(r-r_0)$ and
\begin{equation}\label{alpha}
\alpha(\Theta)  \equiv \left(1-\frac{1}{3}\left[\sqrt{\Omega_{\rm
v}}+(\sqrt{\Omega_{\rm w}} - \sqrt{\Omega_{\rm
v}})\Theta\right]\right)^2  \left(1-\left[\sqrt{\Omega_{\rm
v}}+(\sqrt{\Omega_{\rm w}} - \sqrt{\Omega_{\rm
v}})\Theta\right]^2\right)~.
\end{equation}
To calculate the backreaction (\ref{backreaction t0}), we thus
need to evaluate $v$, $[H_0]$, $[H_0^2]$, $[r H_0']$, and $[r H_0'
H_0]$.

An integral appearing often in these quantities is
\begin{equation}\label{I}
\mathcal{I}(\mu,\nu,\alpha) \equiv \int_\mu^\nu\frac{y^2 {\rm
d}y}{\sqrt{1+y^2\alpha}}~,
\end{equation}
which can be calculated analytically to yield
\begin{equation}\label{Integrated}
\mathcal{I}(\mu,\nu,\alpha) = \frac{1}{2 \alpha^{3/2} } \left[
\sqrt{\alpha} ( \nu \sqrt{1+\alpha \nu^2} - \mu \sqrt{1+\alpha
\mu^2 }) + \ln  \left(\frac{ \sqrt{\alpha} \mu+ \sqrt{1+\alpha
\mu^2 } }{ \sqrt{\alpha}\nu+ \sqrt{1+\alpha \nu^2}} \right)
\right]~.
\end{equation}
For example the volume (\ref{Volume2}) can be written in terms of
the function (\ref{I}) as
\begin{equation}\label{v}
v = \int_0^{\varepsilon}\frac{y^2{\rm d}y}{\sqrt{1+y^2\alpha(0)}}
+ \int_{\varepsilon}^{\varepsilon x}\frac{y^2{\rm
d}y}{\sqrt{1+y^2\alpha(1)}} =
\underbrace{\mathcal{I}(0,\varepsilon,\alpha(0))}_{\equiv
\mathcal{I}_0} + \underbrace{\mathcal{I}(\varepsilon,\varepsilon
x, \alpha(1))}_{\equiv \mathcal{I}_1} ~,
\end{equation}
where $\alpha$ is defined in Eq.\ (\ref{alpha}). Using the
definitions (\ref{I}) and (\ref{v}), $[H_0]$ and $[H_0^2]$ can be
integrated to yield
\begin{eqnarray}\label{H0}
[H_0] &=& t_0^2\left\{\left(1-\frac{\sqrt{\Omega_{\rm
v}}}{3}\right)\mathcal{I}_0+\left(1-\frac{\sqrt{\Omega_{\rm
w}}}{3}\right)\mathcal{I}_1\right\} \\
\label{H02} [H_0^2] &=& t_0\left\{\left(1-\frac{\sqrt{\Omega_{\rm
v}}}{3}\right)^2\mathcal{I}_0+\left(1-\frac{\sqrt{\Omega_{\rm
w}}}{3}\right)^2\mathcal{I}_1\right\}~,
\end{eqnarray}
which combine to give
\begin{equation}\label{firstterminQ}
\left\langle H_0^2 \right\rangle - \left\langle H_0
\right\rangle^2 = \frac{1}{9} \frac{(\sqrt{\Omega_{\rm w}}
-\sqrt{\Omega_{\rm v}})^2}{v^2} \mathcal{I}_0 \mathcal{I}_1 ~.
\end{equation}
Similarly, using Eq.\ (\ref{H0prime}), we have for the derivative
terms
\begin{eqnarray}\label{rH0prime}
[rH_0'] = &-& \frac{t_0^3}{3} (\sqrt{\Omega_{\rm w}}
-\sqrt{\Omega_{\rm v}}) \int_0^{\varepsilon
x}\frac{y^3\delta(t_0y-r_0){\rm d}y}{\sqrt{1+y^2\alpha(\Theta)}}
\\
\nonumber [rH_0H_0'] = &-& \frac{t_0}{3} \varepsilon^3
(\sqrt{\Omega_{\rm w}} -\sqrt{\Omega_{\rm v}})
\left\{\left(1-\frac{\sqrt{\Omega_{\rm
v}}}{3}\right)\int_0^{\varepsilon x}\frac{y^3\delta(t_0y-r_0){\rm
d}y}{\sqrt{1+y^2\alpha(\Theta)}}+ \right.\\
\label{rH0H0prime} &-& \left. \frac{1}{3} (\sqrt{\Omega_{\rm w}}
-\sqrt{\Omega_{\rm v}}) \int_0^{\varepsilon x}\frac{y^3 \Theta(t_0
y -r_0) \delta(t_0y-r_0){\rm d}y}{\sqrt{1+y^2\alpha(\Theta)}}
\right\}~.
\end{eqnarray}
The integrals in Eqs.\ (\ref{rH0prime}) and (\ref{rH0H0prime}) can
be recognized as Stieltjes integrals: using
\begin{equation}\label{changeofvar}
\frac{{\rm d}\Theta(t_0y-r_0)}{{\rm d}y} = t_0\delta(t_0y-r_0)~,
\end{equation}
they can be written as the following ordinary integrals
\begin{eqnarray}\label{A}
\mathcal{A}  &\equiv& \int_0^1\frac{{\rm
d}\vartheta}{\sqrt{1+\varepsilon^2\alpha(\vartheta)}} \\
 \label{B} \mathcal{B}
&\equiv& \int_0^1\frac{\vartheta {\rm
d}\vartheta}{\sqrt{1+\varepsilon^2\alpha(\vartheta)}}~,
\end{eqnarray}
so that Eqs.\ (\ref{rH0prime}) and (\ref{rH0H0prime}) become
\begin{eqnarray}\label{rH0prime2}
[rH_0'] &=& -\frac{t_0^2}{3} \left( \sqrt{\Omega_{\rm w}}
-\sqrt{\Omega_{\rm v}} \right) \mathcal{A} \\
\label{rH0H0prime2} [rH_0H_0'] &=& -\frac{t_0}{3} \varepsilon^3
\left( \sqrt{\Omega_{\rm w}} -\sqrt{\Omega_{\rm v}} \right)
\left\{\left(1-\frac{\sqrt{\Omega_{\rm v}}}{3}\right)\mathcal{A}
-\frac{1}{3} \left(\sqrt{\Omega_{\rm w}} -\sqrt{\Omega_{\rm
v}}\right) \mathcal{B} \right\}.
\end{eqnarray}

By recalling the definition (\ref{average}) and substituting the
expressions (\ref{H0}), (\ref{firstterminQ}), (\ref{rH0prime2})
and (\ref{rH0H0prime2}) in Eq.\ (\ref{backreaction t0}), we
finally obtain:
\begin{equation}\label{BackreactionFinal}
\mathcal{Q} = t_0^{-2}\frac{(\sqrt{\Omega_{\rm w}} -
\sqrt{\Omega_{\rm
v}})^2}{v^2}\left\{\frac{2}{3}\mathcal{I}_0\mathcal{I}_1 +
\frac{4}{9}\big[ \left(\mathcal{I}_0 +
\mathcal{I}_1\right)\mathcal{B} - \mathcal{I}_1\mathcal{A}\big]
\varepsilon^3 - \frac{2}{27} \mathcal{A}^2 \varepsilon^6
\right\}~,
\end{equation}
where the integrals (\ref{A}) and (\ref{B}) can be calculated
numerically or as an expansion in powers of $\varepsilon$ or
$\varepsilon^{-1}$.

\subsection{Power expansions and a fitting formula}\label{powerexpansionofQ}

As the cosmological observations suggest that voids in the cosmic
matter distribution satisfy $r_0 \ll t_0$
\cite{Hoyle:2003hc,Gurzadyan:2008yx,Hunt:2008wp,Gurzadyan:2008va,Gurzadyan:2009if},
one of the most useful methods to investigate the backreaction is
the series expansion in powers of $\varepsilon = r_0/t_0$. By
expanding the exact result (\ref{BackreactionFinal}) in powers of
$\varepsilon$ to second order, we obtain:
\begin{eqnarray}\label{BackreactionExp}
\mathcal{Q} &=& \frac{1}{405}t_0^{-2}\left(\frac{r_0}{R}\right)^3
(\Omega_{{\rm w}}^{1/2} - \Omega_{{\rm v}}^{1/2})^2 \left(54 -
\Omega_{\rm w}^2 - 6 \Omega_{\rm v}^2 - 15 \Omega_{\rm w}
\Omega_{\rm v} - 6 \Omega_{\rm w}^{3/2} \Omega_{{\rm v}}^{1/2} +
\right.
\nonumber \\
&-& \left. 20 \Omega_{\rm v}^{3/2} \Omega_{{\rm w}}^{1/2} + 45
\Omega_{\rm w} \Omega_{{\rm v}}^{1/2} + 90 \Omega_{\rm v}
\Omega_{{\rm w}}^{1/2} - 80 \Omega_{{\rm v}}^{1/2} \Omega_{{\rm
w}}^{1/2} + 36 \Omega_{\rm v}^{3/2} +\right.
\nonumber \\
&+& \left. 9 \Omega_{\rm w}^{3/2} - 20 \Omega_{\rm w} - 48
\Omega_{\rm v} - 30 \Omega_{{\rm w}}^{1/2} - 36 \Omega_{{\rm
v}}^{1/2} \right) \left(\frac{r_0}{t_0}\right)^2 + \mathcal{O}
\left( \frac{r_0}{t_0} \right)^4~,
\end{eqnarray}
while for the values $\Omega_{{\rm v}}=0$ and $\Omega_{{\rm w}}=1$
the fourth order expansion reads as
\begin{equation}\label{BackreactionExp4}
\mathcal{Q} = t_0^{-2} \left(\frac{r_0}{R} \right)^3 \left\{
\frac{4}{135} \left(\frac{r_0}{t_0}\right)^2 -
 \left[ \frac{2677}{102060} -  \frac{3359}{437400}
\left(\frac{r_0}{R} \right)^3 \right]
\left(\frac{r_0}{t_0}\right)^4 \right\} + \mathcal{O} \left(
\frac{r_0}{t_0} \right)^6 ~,
\end{equation}
where the first term in the expansion agrees with the result given
by Eq.\ (5.11) in paper I. For the values $\Omega_{{\rm v}}=0$,
$\Omega_{{\rm w}}=1$ and $(r_0/R)^3=1/2$, the sixth order
expansion of Eq.\ (\ref{BackreactionFinal}) yields:
\begin{equation}\label{BackreactionExp6}
\mathcal{Q} = \frac{2}{135}
\left(\frac{r_0}{t_0}\right)^2-\frac{137107}{12247200}
\left(\frac{r_0}{t_0}\right)^4+\frac{33336241}{3940536600}
\left(\frac{r_0}{t_0}\right)^6 - \mathcal{O} \left(
\frac{r_0}{t_0} \right)^8~.
\end{equation}

\begin{figure}[tbh]
\begin{center}
\includegraphics[width=7.5cm]{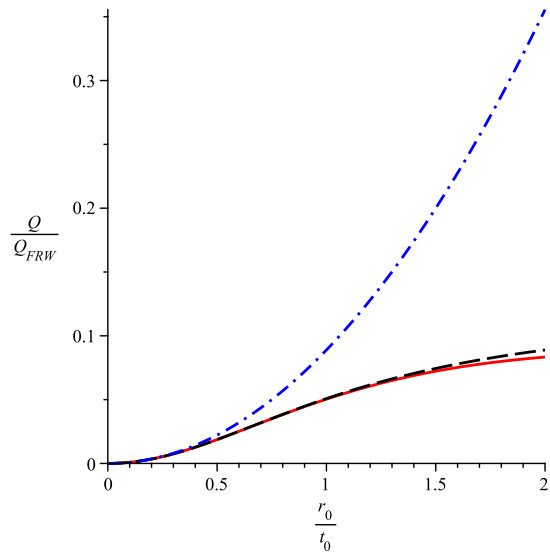}
\caption{The backreaction as a function of $r_0/t_0$ for the
profile (\protect \ref{LTBomegaspecified}) with $\Omega_{\rm v} =
0$, $\Omega_{\rm w} = 1$ and $(r_0/R)^3 = 1/2$ calculated using
the exact result (\protect \ref{BackreactionFinal}) (red solid
curve), the leading order term of the expansion (\protect
\ref{BackreactionExp6}) (blue dash dotted curve) and the fitting
formula (\protect \ref{BackreactionInterpolation}) (black dashed
curve).} \label{Qfit}
\end{center}
\end{figure}

By inspecting the power series (\ref{BackreactionExp6}), we see
that a very simple but accurate fitting formula for the
backreaction in terms of elementary functions is given by
\begin{equation}\label{BackreactionInterpolation}
\mathcal{Q} = \frac{2}{135} t_0^{-2} \left( \frac{r_0}{t_0}
\right)^2 \frac{1}{\displaystyle 1 + \frac{3}{4}\left(
\frac{r_0}{t_0} \right)^2}~.
\end{equation}
Testing the fitting formula (\ref{BackreactionInterpolation})
numerically shows that it is very accurate up to horizon sized
voids $r_0 \sim t_0$ (for $r_0<t_0$ we have $|{\rm error}|<1\%$)
and an excellent approximation even beyond. To illustrate this, we
have plotted the fitting formula (\ref{BackreactionInterpolation})
against the exact backreaction (\ref{BackreactionFinal}) and the
leading order term of the expansion (\ref{BackreactionExp6}) in
Figure \ref{Qfit}. The figure also shows that the mere leading
order term gives an accurate approximation for the backreaction
even up to voids of size $r_0 \sim t_0/3$.

\begin{figure}[tbh]
\begin{center}
\includegraphics[width=7.5cm]{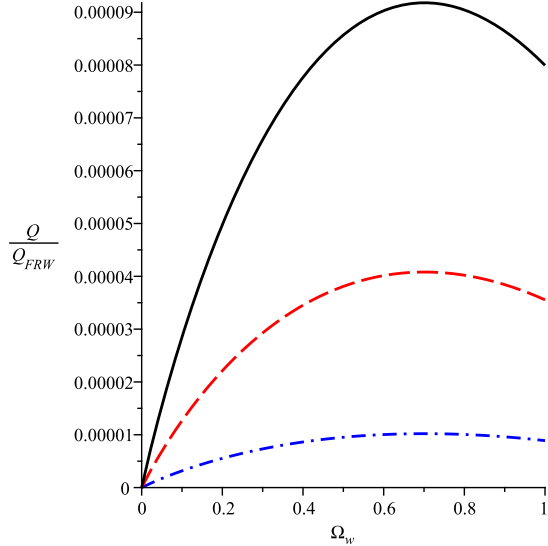}
\caption{The backreaction as a function of $\Omega_{\rm w}$ for
the profile (\protect \ref{LTBomegaspecified}) with $r_0/t_0=
0.01$ (blue dash dotted curve), $r_0/t_0=0.02$ (red dashed curve)
and $r_0/t_0=0.03$ (black solid curve). All profiles in this
figure have $\Omega_{\rm v} = 0$ and $(r_0/R)^3 = 1/2$.}
\label{Q0}
\end{center}
\end{figure}

By solving for the density parameters $\Omega_{{\rm v}}$ and
$\Omega_{{\rm w}}$ that yield the maximum value for the
backreaction of small voids (\ref{BackreactionExp}), we obtain
$\Omega_{{\rm v}}=0$ and $\Omega_{{\rm w}}=0.7$. This can be seen
in Figure \ref{Q0}, where the backreaction is plotted as a
function of $\Omega_{{\rm w}}$ for three values of $r_0/t_0$ and
fixed $\Omega_{\rm v} = 0$ and $(r_0/R)^3 = 1/2$. The figure also
shows that the backreaction is only slightly larger for
$\Omega_{{\rm w}}=0.7$ than for $\Omega_{{\rm w}}=1$.

\subsection{Network of voids}\label{voidnetwork}

Using LTB solutions with the profile defined by Eqs.\
(\ref{LTBomegaspecified}) and (\ref{HoLTB}), we can join together
many void-wall pairs to construct a model for a network of voids
with different $\Omega_{\rm v}$ and $r_0$, presuming the following
matching conditions are met: the different void-wall pairs must
have the same wall-density $\Omega_{\rm w}$ and the same age of
the universe $t_0$. This provides an exact solution, because,
outside the void at $R>r_0$, the void-wall LTB profiles are
identical to the homogeneous Friedmann dust solution with
$\Omega_{\rm w}$ as the density parameter and $t_0$ as the age of
the universe and thus naturally match together.

For a network of voids, the total backreaction $\mathcal{Q}_{{\rm
tot}}$ is given by the formula (\ref{TotalBackreaction}): the
volume-weighted average of the backreactions $\mathcal{Q}_i$ of
the individual void-wall pairs $i$ plus a sum over the relative
variances of the expansion rates between all the different pairs.
The relative variance term makes the configuration particularly
interesting as it seems to offer a way to increase the variance of
the expansion rate without having to introduce any
counterbalancing extra shear.

Let us determine an upper limit to the relative variance term for
a network of voids. As the relative variance is given by the
double sum term in the expression (\ref{TotalBackreaction}), we
need to calculate the volume-average expansion $\langle
\theta\rangle$ for a void-wall LTB model as a function of the
parameters $\Omega_{{\rm v}}$, $\Omega_{{\rm w}}$, $r_0$, $t_0$
and $R$:
\begin{eqnarray}\label{ThetaLTBvoidwallOmega}
\langle \theta\rangle &=& (3-\sqrt{\Omega_{{\rm w}}})t_0^{-1}
+\frac{1}{540}t_0^{-1}\left(\frac{r_0}{R}\right)^3
(\sqrt{\Omega_{{\rm w}}} -\sqrt{\Omega_{{\rm v}}} ) \Big( 108 +
12 \Omega_{{\rm v}}^2 - 6 \Omega_{{\rm v}}^{3/2} \sqrt{\Omega_{{\rm w}}} +  \nonumber \\
&+& 18 \sqrt{\Omega_{{\rm v}}} - 90 \sqrt{\Omega_{{\rm w}}} + 45
\sqrt{\Omega_{{\rm v}}} \Omega_{{\rm w}} + 45 \sqrt{\Omega_{{\rm
w}}} \Omega_{{\rm v}} - 80 \Omega_{{\rm w}} - 80 \sqrt{
\Omega_{{\rm v}} \Omega_{{\rm w}}  } + 64
\Omega_{{\rm v}} +  \nonumber \\
 &-& 63 \Omega_{{\rm v}}^{3/2} - 6 \sqrt{
\Omega_{{\rm v}}} \Omega_{{\rm w}}^{3/2} - 6 \Omega_{{\rm w}}
\Omega_{{\rm v}} + 45 \Omega_{{\rm w}}^{3/2} - 6 \Omega_{{\rm
w}}^2 \Big)\left(\frac{r_0}{t_0}\right)^2~.
\end{eqnarray}

Given the matching condition $\Omega_{{\rm w}}^{(i)}=\Omega_{{\rm
w}}^{(j)}$ between different void-wall LTB regions $i$ and $j$,
the leading order terms of the volume expansion
(\ref{ThetaLTBvoidwallOmega}) cancel in the difference $\langle
\theta\rangle_i - \langle \theta\rangle_j$. Consequently, the
relative variance term in Eq.\ (\ref{TotalBackreaction}) gets the
maximum value when $\Omega_{{\rm v}}^{(i)} = 0$ and $\Omega_{{\rm
v}}^{(j)}=1$, yielding the upper limit:
\begin{equation}\label{relativeVarianceMax}
(\langle \theta\rangle_i - \langle \theta\rangle_j)^2 \leq \left[
\frac{23}{540} t_0^{-1} \left( \frac{r_0}{t_0} \right)^2
\right]^2~.
\end{equation}
Therefore we have
\begin{equation}\label{relativeVarianceMax2}
\frac{1}{3}\sum_{i=1}^{N}\sum_{j=1}^{N}
f_if_j\left(\langle\theta\rangle_{\mathcal{D}_i}-\langle\theta\rangle_{\mathcal{D}_j}\right)^2
\leq \frac{1}{3} \sum_{i=1}^{N}\sum_{j=1}^{N} \frac{1}{N^2}
\frac{529}{291600} t_0^{-2} \left( \frac{r_0}{t_0} \right)^4
\end{equation}
or
\begin{equation}\label{relativeVarianceMax3}
\frac{1}{3}\sum_{i=1}^{N}\sum_{j=1}^{N}
f_if_j\left(\langle\theta\rangle_{\mathcal{D}_i}-\langle\theta\rangle_{\mathcal{D}_j}\right)^2
\leq 10^{-3} t_0^{-2} \left( \frac{r_0}{t_0} \right)^4~,
\end{equation}
where $r_0$ is the radius of the largest void in the network. Eq.\
(\ref{relativeVarianceMax3}) shows that the relative variance term
is at most of the order $\left( r_0/t_0 \right)^4$ so the
backreaction for a network of LTB voids with profile defined by
Eqs.\ (\ref{LTBomegaspecified}) and (\ref{HoLTB}) is essentially
no greater than the backreaction for a single void-wall pair.

\subsection{Large voids}\label{largevoids}

Let us then study the backreaction of large voids, i.e.\ voids
that do not satisfy $r_0 \ll t_0$. For voids of the horizon size
$r_0 \sim t_0$, the exact backreaction (\ref{BackreactionFinal})
has to be evaluated numerically or using the fitting formula
(\ref{BackreactionInterpolation}) but for superhorizon voids with
$r_0 \gg t_0$, we can consider the limit $r_0/t_0 \rightarrow
\infty$ of Eq.\ (\ref{BackreactionFinal}), which, after some
manipulations, takes the form:
\begin{eqnarray}\label{Qsuperhorizon} \nonumber
\mathcal{Q}_{\infty} = t_0^{-2} \frac{\alpha_{{\rm v}}\alpha_{{\rm
w}}(\sqrt{\Omega_{{\rm w}}}-\sqrt{\Omega_{{\rm v}}})^2}{(
\sqrt{\alpha_{{\rm w}}} + \sqrt{\alpha_{{\rm v}}}(x^2-1) )^2}
\left[ \frac{2}{3} \frac{(x^2-1)}{\sqrt{\alpha_{{\rm v}}
\alpha_{{\rm w}}}} - \frac{8}{27} \left(\int_{0}^{1} \frac{{\rm d}
\vartheta}{\sqrt{\alpha(\vartheta)}} \right)^2 + \right. \\
\left.  + \frac{8}{9} \frac{\sqrt{\alpha_{{\rm
w}}}+\sqrt{\alpha_{{\rm v}}}(x^2-1)}{\sqrt{\alpha_{{\rm
v}}\alpha_{{\rm w}}}} \int_{0}^{1} \frac{\vartheta {\rm d}
\vartheta}{\sqrt{\alpha(\vartheta)}} - \frac{8}{9}
\frac{(x^2-1)}{\sqrt{\alpha_{{\rm w}}}} \int_{0}^{1} \frac{ {\rm
d} \vartheta}{\sqrt{\alpha(\vartheta)}} \right]~,
\end{eqnarray}
where $\alpha(\vartheta)$ is defined in Eq.\ (\ref{alpha}),
$\alpha_{{\rm v}} \equiv \alpha(0)$, $\alpha_{{\rm w}} \equiv
\alpha(1)$ and the integrals can be evaluated in terms of
elementary functions:
\begin{equation}\label{Alimit}
\int_{0}^{1} \frac{ {\rm d} \vartheta}{\sqrt{\alpha(\vartheta)}} =
\left. \frac{3 \sqrt{2}}{4(\sqrt{\Omega_{{\rm
w}}}-\sqrt{\Omega_{{\rm v}}})} \arctan \left( \frac{3 \sqrt{2}}{4}
\frac{ (\vartheta - \frac{1}{3}) }{\sqrt{1 - \vartheta^2 }}
\right) \right|_{\sqrt{\Omega_{{\rm v}}}}^{\sqrt{\Omega_{{\rm
w}}}} ~,
\end{equation}
\begin{eqnarray}\label{Blimit} \nonumber
\int_{0}^{1} \frac{ \vartheta {\rm d}
\vartheta}{\sqrt{\alpha(\vartheta)}} &=&
\frac{1}{(\sqrt{\Omega_{{\rm w}}}-\sqrt{\Omega_{{\rm v}}})^2}
\Biggr\{ 3 \left(\arcsin \sqrt{\Omega_{{\rm v}}} - \arcsin \sqrt{\Omega_{{\rm w}}} \right)   \\
&+& \frac{3 \sqrt{2}}{4} (3 - \sqrt{\Omega_{{\rm v}}}) \left.
\arctan \left( \frac{3 \sqrt{2}}{4} \frac{ (\vartheta -
\frac{1}{3}) }{\sqrt{1 - \vartheta^2 }} \right)
\right|_{\sqrt{\Omega_{{\rm v}}}}^{\sqrt{\Omega_{{\rm w}}}}
\Biggr\}~.
\end{eqnarray}

\begin{figure}[tbh]
\begin{center}
\includegraphics[width=7.5cm]{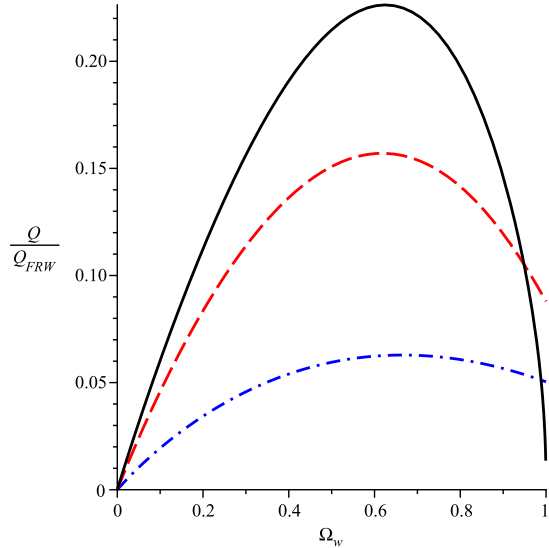}
\caption{The backreaction as a function of $\Omega_{\rm w}$ for
the profile (\protect \ref{LTBomegaspecified}) with $r_0/t_0= 1$
(blue dash dotted curve), $r_0/t_0=3$ (red dashed curve) and
$r_0/t_0 \rightarrow \infty$ (black solid curve). All profiles in
this figure have $\Omega_{\rm v} = 0$ and $(r_0/R)^3 = 1/2$.}
\label{Q3}
\end{center}
\end{figure}

Eq.\ (\ref{Qsuperhorizon}) can be used to solve the values of the
density parameters $\Omega_{{\rm v}}$ and $\Omega_{{\rm w}}$ that
maximize the backreaction for the superhorizon voids: numerically,
we find that the maximum is located at $\Omega_{{\rm v}}=0$ and
$\Omega_{{\rm w}}=0.625$ and has the value
\begin{equation}\label{MaxBackreaction}
\mathcal{Q}_{{\rm max}} = 0.038 t_0^{-2} = 0.23 \mathcal{Q}_{{\rm
FRW}} ~,
\end{equation}
which can also be read off from Figure \ref{Q3}. The location of
the maximum differs only slightly from the point $\Omega_{{\rm
v}}=0~$, $\Omega_{{\rm w}}=0.7$ that gives the maximum
backreaction for small $r_0 \lesssim 0.3 \hspace{2pt} t_0$ voids,
although the maximum value of the backreaction relative to the
value at $\Omega_{{\rm w}}=1$ is in this case noticeably larger
(cf.\ Sect.\ \ref{powerexpansionofQ}). However, even with the
superhorizon sized voids, the maximum backreaction
(\ref{MaxBackreaction}) is still only $\sim 20\%$ of the FRW value
(\ref{QFRW}), implying that in the void-wall LTB models with the
profile defined by Eqs.\ (\ref{LTBomegaspecified}) and
(\ref{HoLTB}) the shear suppresses the backreaction in all cases
at least by about an order of magnitude.

\begin{figure}[tbh]
\begin{center}
\includegraphics[width=8.0cm]{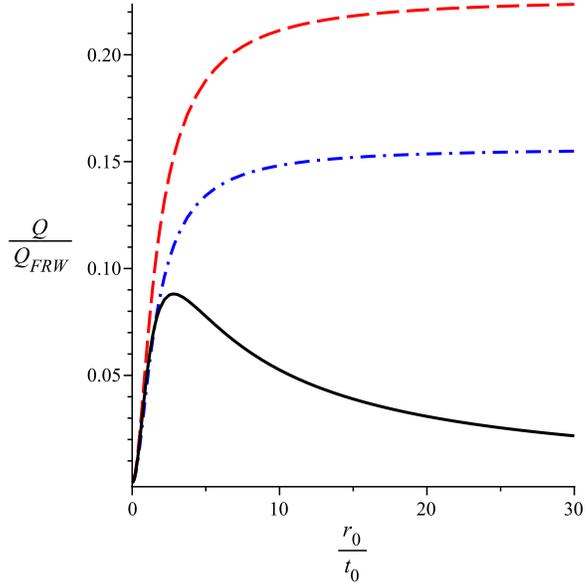}
\caption{The backreaction as a function of $r_0/t_0$ for the
profile (\protect \ref{LTBomegaspecified}) with $\Omega_{\rm w} =
1$ (black solid curve), $\Omega_{\rm w} = 0.3$ (blue dash dotted
curve) and $\Omega_{\rm w} = 0.62$ (red dashed curve). All
profiles in this figure have $\Omega_{\rm v} = 0$ and $(r_0/R)^3 =
1/2$.} \label{Q1}
\end{center}
\end{figure}

In Figure \ref{Q1}, we have plotted the backreaction as a function
of the size of the void $r_0/t_0$ for different values of the wall
density $\Omega_{\rm w}$ but keeping the void density and the
volume fraction fixed: $\Omega_{\rm v} = 0$ and $(r_0/R)^3 = 1/2$.
The figure illustrates how the rapid growth of the backreaction as
a function of the void size $r_0$ stops once the horizon size
$t_0$ is exceeded.


\subsection{Uncompensated voids}\label{kompensoimatonna}

As the density function $\Omega_0(r)$ is determined by the
integral of the physical matter density (\ref{LTBomega}), the
profile (\ref{LTBomegaspecified}) implies an overdense shell in
$\rho_0(r)$ between the void and the wall (see Sect.\ 5.4 of paper
I for a more detailed explanation). To avoid the overdense or
collapsing shell, we consider the profile defined by
\begin{equation}\label{LTBuncompensated}
\theta(r,t_0) = t_0^{-1} \left( 3 - \Theta(r-r_0) \right)~,
\end{equation}
which, by virtue of Eqs.\ (\ref{H0-OmegaRelationAppr}) and
(\ref{LTBtheta}), implies
\begin{equation}\label{omegathetastep}
\Omega_0(r) = \left( 1 - \left(\frac{r_0}{r}\right)^3 \right)^2
\Theta(r-r_0)
\end{equation}
and
\begin{equation}\label{H0thetastep}
H_0(r) = t_0^{-1} \left[1 -  \frac{1}{3} \left( 1 - \left(
\frac{r_0}{r} \right)^3 \right) \Theta (r-r_0) \right]~.
\end{equation}

\begin{figure}[tbh]
\begin{center}
\includegraphics[width=7.5cm]{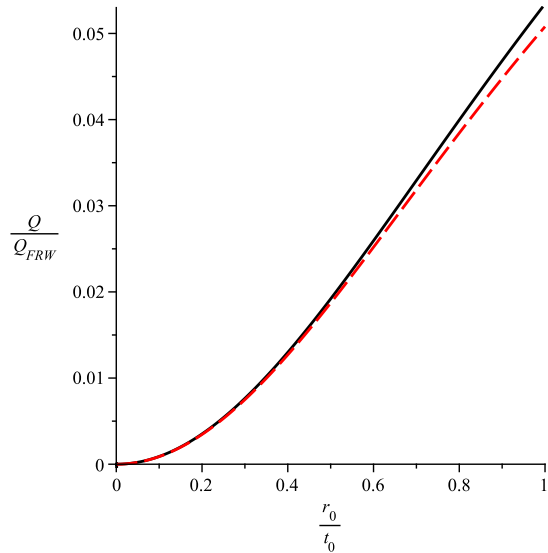}
\caption{The backreaction as a function of $r_0/t_0$ for the
compensated void-wall profile (\protect \ref{LTBomegaspecified})
with $\Omega_{\rm v}=0$ and $\Omega_{\rm w}=1$ (red dashed curve)
and for the uncompensated void-wall profile (\protect
\ref{LTBuncompensated}) (black solid curve). Both profiles in this
figure have $(r_0/R)^3 = 1/2$.} \label{uncomp}
\end{center}
\end{figure}

In paper I, we found that the difference in the backreaction
between the compensated and the uncompensated void was
$\mathcal{O}(10\%)$ for small voids. In Figure \ref{uncomp}, we
have plotted the backreaction for the profile defined by Eqs.\
(\ref{omegathetastep}) and (\ref{H0thetastep}) to show that the
difference remains small even for large voids. This proves that
the suppression of the backreaction is not due to the compensating
overdensity but truly an effect of the shear.

\section{Summary of the results}\label{konkluusiot}

We have generalized the analytic approach of Ref.\
\cite{Mattsson:2010vq} or paper I in studying the role of shear in
the cosmological backreaction problem. As in paper I, we
constructed a void-wall model from the exact inhomogeneous LTB
dust solution, but instead of assuming the voids to be small
$r_0\ll t_0$ and fixing the densities in the voids and walls to
$\Omega_{\rm{v}}=0$ and $\Omega_{\rm{w}}=1$, we let the void size
$r_0$ be arbitrary and the densities vary within the range $0 \leq
\Omega_{\rm{v}}\leq\Omega_{\rm{w}}\leq 1$. Moreover, we
constructed an exact solution for a network of voids with
different densities $\Omega_{\rm{v}}$ and radii $r_0$ to study how
the emerging relative variance of the expansion rate between the
different void-wall pairs affects the backreaction.

We calculated the exact analytic result for the backreaction of a
void-wall pair with profile defined by Eqs.\
(\ref{LTBomegaspecified}) and (\ref{HoLTB}) in Sect.\
\ref{generalQ}. To study its general behavior for subhorizon
voids, we considered the series expansion in powers of
$r_{0}/t_{0}$ in Sect.\ \ref{powerexpansionofQ}. The leading order
term of order $(r_0/t_0)^2$ in the expansion was verified to agree
with the result obtained in paper I for the case
$\Omega_{\rm{v}}=0$ and $\Omega_{\rm{w}}=1$. From the leading
order term we found that the values $\Omega_{\rm{v}}=0$ and
$\Omega_{\rm{w}}=0.7$ yield the maximum value for the backreaction
for subhorizon voids of size $r_0\lesssim t_0/3$. However, the
increase in the backreaction was showed to be only $\sim 10 \%$
relative to the case where $\Omega_{\rm{v}}=0$ and
$\Omega_{\rm{w}}=1$. Furthermore, by inspecting the series
expansion with the values $\Omega_{\rm{v}}=0$, $\Omega_{\rm{w}}=1$
and $(r_{0}/R)^{3}=1/2$, we were able to deduce the simple fitting
formula (\ref{BackreactionInterpolation}) for the backreaction in
terms of elementary functions, with error less than $1\%$ up to
horizon sized voids.

In Sect.\ \ref{voidnetwork}, we considered a network of subhorizon
voids and showed that the relative variance of the expansion rate
between the different voids is at most of order $(r_0/t_0)^4$. The
total backreaction for the network is then essentially given just
by the volume-weighted average of the backreactions of the
individual void-wall pairs and thus remains of order $(r_0/t_0)^2$
or small for voids of the observed size.

We investigated the behavior of the backreaction for large voids
in Sect.\ \ref{largevoids}, in particular superhorizon $r_0\gg
t_0$ voids by considering the limit $r_0/t_0 \rightarrow \infty$.
We found that the backreaction approaches a constant value as the
size of a superhorizon void is increased, the limiting value
depending on the parameters $\Omega_{\rm{v}}$, $\Omega_{\rm{w}}$
and $R/r_0$. The ultimate maximum value for the backreaction was
found to be $\mathcal{Q}_{\rm{max}} = 0.038t_0^{-2}$ at
$\Omega_{\rm{v}}=0$ and $\Omega_{\rm{w}}=0.62$ in the limit
$r_0/t_0 \rightarrow \infty$, keeping the void-to-wall volume
ratio fixed to $(r_{0}/R)^{3}=1/2$. This is still only $\sim 20
\%$ of the maximum backreaction obtained when the shear is
neglected by treating the voids and walls as separate Friedmann
solutions. Furthermore, as opposed to the $\sim 10 \%$ increase in
the case of small voids, the backreaction for large voids is a few
times (depending on the ratio $r_0/t_0$) greater at the maximum
point $\Omega_{\rm{v}}=0$ and $\Omega_{\rm{w}}=0.62$ than at
$\Omega_{\rm{v}}=0$ and $\Omega_{\rm{w}}=1$. Finally, in Sect.\
\ref{kompensoimatonna} we demonstrated that the suppression of the
backreaction is not due to an overdense or collapsing region
between the void and the wall but truly an effect of the shear.

\section{Conclusion}\label{konkluusiot2}

By considering exact solutions of the Einstein equation consisting
of one or more LTB solutions, we have pinpointed the issue of
small versus large cosmological backreaction to the question of
matching conditions: while the variance of the expansion rate
alone can induce significant backreaction, the shear arising from
matching together the regions with different expansion rates seems
to bring down the backreaction by at least five orders of
magnitude for voids of the observed size. The crucial question is
whether the suppression of the backreaction due to the shear is a
general property of all realistic cosmological solutions of
general relativity or just a special property of the matching in
the considered particular solutions. This issue has to be
addressed with solutions more sophisticated than the LTB-based
models employed here.

\acknowledgments{We thank David Wiltshire, Krzysztof Bolejko,
Syksy R\"as\"anen, Kari Enqvist, Kimmo Kainulainen, Pyry Rahkila
and Valerio Marra for helpful comments and discussions. This work
was supported by the Marsden fund of the Royal Society of New
Zealand. MM is supported by the Graduate School in Particle and
Nuclear Physics (GRASPANP) and acknowledges The Magnus Ehrnrooth
foundation for supporting her visit in the University of
Canterbury and David Wiltshire for hospitality during her stay in
Canterbury.}

\end{document}